\documentclass[reprint,showpacs,amsmath,amssymb,aps]{revtex4}
\usepackage{graphicx,color}
\newcommand{\Array}[2]{\left(\begin{array}{#1}#2\end{array}\right)}

\newcommand\red[1]{\textcolor{red}{#1}}

\begin{document}

\title{The Structure of Flavor Mixing and Reconstruction of the Mass Matrix}

\author{Guojun Xu$^a$, %email:15955124@stu.xjtu.edu.cn
Jingjun Zhang$^a$, %email:zjj2240250379@stu.xjtu.edu.cn
Chenzi Liao$^a$, %email:lcz020108@stu.xjtu.edu.cn
Ying Zhang$^{a,b}$\footnote{E-mail: hepzhy@mail.xjtu.edu.cn. Corresponding author. }}
\address{$^a$School of Physics, Xi'an Jiaotong University, Xi'an, 710049, China}

\address{$^b$Institute of Theoretical Physics, Xi'an Jiaotong University, Xi'an, 710049, China}

\begin{abstract}
The fermion flavor structure is investigated by bilinear decomposition of the mass matrix after EW symmetry breaking, and the roles of factorized matrices in flavor mixing and mass generation are explored.  On a new Yukawa basis, the minimal parameterization of flavor mixing is realized containing two relative phases and two free $SO(2)_L$ rotation angles. It is shown that flavor mixing can be addressed from 4 independent parameters. The validity of the flavor mixing structure is checked in both the lepton and quark sectors. 
Under the decomposition of flavor mixing, fermion mass matrices are reconstructed in the hierarchy limit. A flat mass matrix with all elements equal to 1 arises naturally from the requirement that homology exists between up-type and down-type fermion mass matrices. Some hints of a flat matrix and flavor breaking are also discussed.
\end{abstract}

\maketitle

\section{Motivation}
Although the Standard Model (SM) has been successful in experiments, puzzles of fermion mass hierarchy and origin of flavor mixing as famous unknown problems remain mysteries \cite{Feruglio:2015jfa,ZZX2020PR,2019Zupan,Haba2001PRD,GrossmanTASI2017}. 
In the SM, a quark mass matrix is generated from Yukawa couplings $${\bf M}^q_{ij}=\frac{v_0}{\sqrt{2}}y^q_{ij},~~q=u,d$$
with Higgs VEV $v_0/\sqrt{2}$.
Physical masses are obtained by a bi-unitary transformation $q_{L,R}=({\bf U}^q_{L,R})^\dag q_{L,R}^{(m)}$ to mass eigenstate labelled by superscript $^{(m)}$
	$${\bf U}^q_L{\bf M}^q ({\bf U}^q_R)^\dag={\rm diag}(m^q_1,m^q_2,m^q_3).$$
The CKM mixing matrix appears in the charged current interaction as
	${\bf U}_{CKM}={\bf U}_L^u{{\bf U}_L^d}^\dag$.
Yukawa couplings $y^{u,d}_{ij}$ are responsible for not only quark hierarchal masses but also CKM mixing with nonvanishing CP violation (CPV). 
However, complex $y^{u,d}_{ij}$ includes many more parameters than what could be measured in experiments.
The SM cannot provide any detail about the value $y^{u,d}_{ij}$ or its structure. This applies to lepton as well. Considering the minimal extended SM with three Dirac neutrinos, the lepton mass matrix is generated by the Yukawa coupling $y_{ij}^{l}$
	\begin{eqnarray}
	{\bf M}^l_{ij}=\frac{v_0}{\sqrt{2}}y^l_{ij},~~l=\nu,e.
	\label{eq.I05}
	\end{eqnarray}
Bi-unitary transformation is performed to diagonalize the mass matrix; then, the physical masses are obtained.
	\begin{eqnarray}
	{\bf U}^l_L{\bf M}^l ({\bf U}^l_R)^\dag={\rm diag}(m^l_1,m^l_2,m^l_3)
	\label{eq.I06}
	\end{eqnarray}
In the lepton charged current interaction, a flavor mixing matrix known as PMNS mixing appears:
	${\bf U}_{PMNS}={\bf U}_L^e{{\bf U}_L^\nu}^\dag$.
Despite mixing angle differences numerically, the structure similarity of CKM and PMNS mixing matrixes strongly implies the existence of a common mechanism behind SM Yukawa couplings.
Decoding the flavor structure and finding the underlying organization principles have become important missions of particle physics today. 
Besides explaining these issues from new physics \cite{Georgi1979PLB,Archer2011JHEP, LiPRD2010},
 recently, some novel ideas, such as flavor mixing substructures \cite{2021JHEPsubstructure} and random Yukawa couplings \cite{Gersdorff2017JHEP}, have also been proposed to address flavor structure issues.

Some clues on hidden flavor structure are discovered from the SM itself.
On the theoretical side, Yukawa terms describe non-gauge interactions between the Higgs and chiral fermions. However, the fermion fields are expressed on a gauge basis. 
A natural idea arises whether there is a new basis (called the Yukawa basis in \cite{YZhang202100}) in which Yukawa couplings become real and even show that some organized structure arises.
Fermions in the Yukawa basis are a linear superposition of ones in the gauge states, which put forward a new standpoint to comprehend the origin of CPV from the superposition coefficients between different flavors. In early research, the idea was realized in a minimal flavor structure model (MFS)\cite{YZhang202100,YZhang202101}. It includes 10 free parameters in the lepton/quark sector to correspond to 6 lepton/quark masses, 3 PMNS/CKM mixing angles, and 1 CP violating phase without other redundant unphysical parameters. 
MFS has been proven successful in generating experimental data in both the quark sector and the lepton sector.

On the other hand, we have obtained almost all fermion masses and PMNS/CKM mixing parameters in experiments.
If assuming normal order neutrino masses, there is a common characteristic in the fermion mass distribution $m_1^f\ll m_2^f\ll m_3^f$ for $f=u,d,\nu,e$. Defining the mass hierarchy $h_{ij}^f=m_i^f/m_j^f$, we find that $h_{12}^f,h_{23}^f\ll1$, which could be treated as zero at the leading order approximation, i.e., $h_{12}^f=h_{23}^f=0$ (called the hierarchy limit). We pose the question of what influences the hierarchy limit could bring on flavor mixing. 
In the SM, mass hierarchy and flavor mixing are confused into ambiguous Yukawa couplings. 
It is unclear whether these two independent questions arise from different principles or mechanics. 
If the answer is yes, mass hierarchy and flavor mixings can be addressed from independent mechanisms.
With the help of the Yukawa basis, flavor mixing can be investigated from a clear point of view.
Uncovering the final flavor structure of fermions is a key point in the future. This is the first mission in this paper. 

Due to the flavor mixing matrix arising from up-type and down-type fermion mass matrix diagonalization transformation, it provides some information about the mass matrix that can be used to reconstruct the mass matrix. Another mission is to investigate the homology of mass matrices, i.e. whether up-type and down-type mass matrices come from a common structure. By decomposing the flavor mixing matrix into factorized matrices, we analyze the difference between up-type and down-type mass matrices to seek a common origin of mass.

We start from the lepton sector with the minimal extended normal order Dirac neutrinos. We review the decomposition of the lepton mass matrix and the concept of the Yukawa basis in the next section.
In Section III, the PMNS structure and parameterization are discussed in the hierarchy limit. Lepton flavor mixing is parameterized independently from lepton masses.
By the decomposition of PMNS, lepton mass matrixes are reconstructed in Section IV. The homology of charge lepton and neutrino mass is focused specifically, which leads to a flat matrix. In Section V, we generalize all results to quarks and show their validity. The inspiration for the homology of fermion mass and flavor breaking is discussed in Section VI. A summary is given in the last section.
%%%%%%%%%%%%%%

\section{Flavor Mixing in the Yukawa Basis}
In the minimal extended SM with three Dirac neutrinos, the lepton Yukawa term is written as
	\begin{eqnarray}
		-\mathcal{L}_Y=y^e_{ij}\bar{L}_L^i\Phi{e^{j}_R}+y^\nu_{ij}\bar{L}_L^i\tilde{\Phi}{\nu^{j}_R}+H.c.
		\label{eq.LepYua10}
	\end{eqnarray}
with left-handed doublet $L_L=(\nu_L,e_L)^T$, right-handed charged lepton and neutrino $e_R,\nu_{R}$ and Higgs doublet $\Phi$.
Here, the generation index $i,j=1,2,3$. 
After electroweak symmetry breaking, the lepton mass matrix becomes ${\bf M}^l$, as shown in eq. (\ref{eq.I05}).
Bi-unitary transformations are performed in flavor space:
\begin{eqnarray}
\nu_{L,R}=({\bf U}^\nu_{L,R})^\dag \nu^{(m)}_{L,R},~~~~
e_{L,R}=({\bf U}^e_{L,R})^\dag e^{(m)}_{L,R}
\label{eq.diagonalizetransf}
\end{eqnarray}
Then, ${\bf M}^l$ is diagonalized to physical masses in eq. (\ref{eq.I06}). 
As mentioned above, a complex mass matrix is generated by flavor-dependent Yukawa couplings and has a dual responsibility: to generate physical masses and to provide PMNS mixings.
Inspired by eq. (\ref{eq.I06}), 
${\bf M}^l$ is decomposed into a real ${\bf M}_0^l$ and two unitary ${\bf F}_{L,R}^l$
\begin{eqnarray}
	{\bf M}^l=({\bf F}_L^l)^\dag{\bf M}_0^l{\bf F}_R^l
	\label{eq.II05}
	\end{eqnarray}
Here, real matrix ${\bf M}_0^l$ can be diagonalized to physical masses by an orthogonal transformation ${\bf U}_0^l$
 $${\bf U}_0^l{\bf M}_0^l({\bf U}_0^l)^T={\rm diag}(m_1^l,m_2^l,m_3^l)$$
Thus, ${\bf M}_0^l$ completely encodes lepton physical masses. 

To reveal the role of unitary ${\bf F}_{L,R}^l$ for convenience, we define a new basis, labeled by superscript $^{(Y)}$, as follows
	\begin{eqnarray}
		\nu_{L,R}^{(Y)}={\bf F}_{L,R}^\nu \nu_{L,R},~~~
		e_{L,R}^{(Y)}={\bf F}_{L,R}^e e_{L,R}
	\label{eq.II06}
	\end{eqnarray}
On the new basis, lepton Yukawa term Eq. (\ref{eq.LepYua10}) can be expressed by real couplings, which generates real  mass  terms 
	\begin{eqnarray}
		-\mathcal{L}_Y=\bar{e}_L^{(Y)}{\bf M}_0^l e_R^{(Y)}
		+\bar{\nu}_L^{(Y)}{\bf M}_0^\nu \nu_R^{(Y)}+H.c.
	\end{eqnarray}
However, the weakly charged lepton current interaction receives contributions from  ${\bf F}_L^{e,\nu}$ 
\begin{eqnarray}
	\mathcal{L}_{CC}^l
	&=&-\frac{g}{\sqrt{2}}\bar{e}_L^{(Y)}\gamma^\mu{\bf F}_L^e({\bf F}_L^\nu)^\dag\nu_L^{(Y)} W_\mu^-
			+h.c.
	\nonumber\\
	&=&-\frac{g}{\sqrt{2}}\bar{e}_L^{(m)}\gamma^\mu{\bf U}_{PMNS}\nu_L^{(m)} W_\mu^-
			+h.c.
\end{eqnarray}
where  PMNS mixing matrix is ${\bf U}_{PMNS}={\bf U}_0^e{\bf F}_L^e({\bf F}_L^\nu)^\dag ({\bf U}_0^\nu)^T$.
Complex phases in ${\bf U}_{PMNS}$ completely come from ${\bf F}_L^e({\bf F}_L^\nu)^\dag$ unless ${\bf F}_L^e={\bf F}_L^\nu$. 
Real orthogonal transformation ${\bf U}_0^{e,\nu}$ cannot generate a complex phase that is required by nonvanishing CPV.  It is shown that the mismatching between up-type and down-type fermions is the origin of CP violation.

Considering only two independent phases between three flavors in a family, a minimal parameterized ${\bf F}_L^{l}$ is proposed by MFS \cite{YZhang202101} with only two relative phases:
	\begin{eqnarray}
		{\bf {F}}^\nu_L={\rm diag}(1,e^{i\lambda^\nu_1},e^{i\lambda^\nu_2}),~~
		{\bf {F}}^e_L={\rm diag}(1,1,1),~~~
	\label{eq.II10}
	\end{eqnarray}
The PMNS matrix has the form
	\begin{eqnarray}
	{\bf U}_{PMNS}= {\bf U}_0^e\text{~diag}\left(1,e^{-i\lambda^\nu_1},e^{-i\lambda^\nu_2}\right) ({\bf U}_0^\nu)^T
	\label{eq.II11}
	\end{eqnarray}
The right-handed ${\bf F}_R^{\nu,e}$ represent nonphysical parameters, which are treated as the identity matrix.
On the new basis, the Yukawa terms after EW symmetry breaking can be rewritten as
	\begin{eqnarray}
		-\mathcal{L}_M^l
			&=&\bar{e}_L^{(Y)} {\bf M}_0^e   e_R^{(Y)}
			+\bar{\nu}_L^{(Y)} {\bf M}_0^\nu  \nu_R^{(Y)}+h.c.
	\end{eqnarray}
with real ${\bf M}_0^{\nu,e}$.
%%%%%%
\section{PMNS in the Hierarchy Limit}
SM fermions (with normal order Dirac neutrino) exhibit a mass hierarchical structure. 
On the mass basis, the charged lepton mass matrix has the following structure:
$${\bf M}_{diag}^e={m_3^e}\Array{ccc}{h_{12}^eh_{23}^e && \\ & h_{23}^e & \\ && 1}$$
Experiments have given $h_{12}^e\simeq0.0048$ and $h_{23}^e\simeq0.059$. For normal-order neutrinos, the mass matrix has a similar structure with $h_{12}^\nu\simeq0.012,h_{23}^\nu\simeq0.172$ for initial $m_1^\nu=0.0001$eV.
Considering the hierarchy limit condition in which $h_{ij}^l$ is treated as a perturbation quantity, the mass matrix normalized by the total family mass at the leading order of $h_{ij}^f$ becomes
	\begin{eqnarray*}
	\frac{1}{\sum_im_i^l}{\bf M}_{diag}^l=\Array{ccc}{0 && \\ & 0& \\ && 1}
	\end{eqnarray*}
In this case, only the third-generation lepton is massive. This scenario was also considered in \cite{Weinberg2020}, 
in which generation of the first and second generation masses from radiative corrections is attempted.
We can investigate the flavor structure by eliminating mass hierarchy corrections. 

Obviously, there is a chiral $SO(2)_L^l\times SO(2)_R^l$ symmetry for $l=\nu,e$ in the real subspace of the first and second ones in the Yukawa basis:
	\begin{eqnarray}
		l_{L}^{(m)}\rightarrow{\bf R}^T_3(\theta_L^l)l_{L}^{(m)},~~
		l_{R}^{(m)}\rightarrow{\bf R}^T_3(\theta_R^l)l_{R}^{(m)}
	\end{eqnarray}
in which ${\bf M}_{diag}^l$ is kept invariant:
$${\bf M}_{diag}^l\rightarrow{\bf R}_3(\theta_L^l){\bf M}_{diag}^l{\bf R}_3^T(\theta_R^l)={\bf M}_{diag}^l.$$
Here, ${\bf R}_i(\theta)$ is a plane rotation around the $i$-th axis.
For PMNS mixing, this symmetry gives the freedom of inserting two left-handed rotation parameters $\theta_L^\nu$ and $\theta_L^e$ into the flavor mixing structure:
	\begin{eqnarray}
	{\bf U}_{PMNS}={\bf R}_3(\theta_L^e){\bf U}_0^e{\rm diag}(1, e^{-i\lambda_1^\nu},e^{-i\lambda_2^\nu})({\bf U}_0^\nu)^T{\bf R}_3^T(\theta_L^\nu),
	\label{eq.PMNSstructure01}
	\end{eqnarray}
Without loss of generality, the real orthogonal rotation ${\bf U}_0^l$ is factorized into 3 rotations around 3 fixed axes:
\begin{eqnarray}
{\bf U}_0^l&=&\Array{ccc}{c_{\theta_3^l}& s_{\theta_3^l}& 0\\
		-s_{\theta_3^l} & c_{\theta_3^l} &0 \\
		0 & 0 & 1}
	\Array{ccc}{c_{\theta_2^l}& 0 & s_{\theta_2^l}\\
		0 & 1 & 0 \\
		-s_{\theta_2^l} & 0 & c_{\theta_2^l}}
	\Array{ccc}{1 & 0 & 0 \\
		0 & c_{\theta_1^l}& s_{\theta_1^l}\\
		0 & -s_{\theta_1^l} & c_{\theta_1^l} }
	\\
	&=&{\bf R}_3(\theta_3^l){\bf R}_2(\theta_2^l){\bf R}_1(\theta_1^l).
\end{eqnarray}
Absorbed above ${\bf R}_3$ into ${\bf R}_3(\theta_L^e)$ and   ${\bf R}_3^T(\theta_L^\nu)$ in eq. (\ref{eq.PMNSstructure01}), the PMNS matrix becomes
	\begin{eqnarray}
	{\bf U}_{PMNS}=  {e^{-i\lambda_0^\nu}}{\bf R}_3(\theta_3^e){\bf R}_2(\theta_2^e){\bf R}_1(\theta_1^e)\text{~diag}\left(1,e^{-i\lambda^\nu_1},e^{-i\lambda^\nu_2}\right){\bf R}_1^T(\theta_1^\nu){\bf R}_2^T(\theta_2^\nu){\bf R}_3^T(\theta_3^\nu)
	\label{eq.III07}
	\end{eqnarray}
Here, a global unphysical phase $e^{-i\lambda_0^\nu}$ has been added at the right side. 
A general $3\times3$ unitary matrix includes $3^2$ d.o.f. (including a global phase).  On a mass basis, each field can be re-defined by a global $U(1)$ random global phase $\psi_i^q\rightarrow e^{i\kappa_i^q}\psi_i^q$. Using the transformation, five unphysical phases can be eliminated, leaving only 3 angles and 1 phase, which corresponds to the same number of observables in PMNS mixings. The process is called rephasing \cite{Branco2012RMP}.
At the right of eq. (\ref{eq.III07}), there are also 9 parameters: 1 global phase $e^{-i\lambda_0^\nu}$,
$3$ charged lepton rotation angles $\theta^l_i$, $3$ neutrino rotation angles $\theta^\nu_i$ and 2 relative phases. After the rephasing process, which eliminates 5 unphysical parameters, only 4 physical quantities remain.

If the mass hierarchy and flavor mixings are two independent questions, the necessary condition is that flavor mixing must be parameterized by 4 quantities that do not depend on the form of the mass matrix.
From the diagonal mass matrix, ${\bf U}_0^l$ can be used to reconstruct the mass matrix
	\begin{eqnarray}
		\frac{1}{\sum_im_i^l}{\bf M}_0^l&=&({\bf U}_0^l)^T\Array{ccc}{0 && \\ & 0& \\ && 1}{\bf U}_0^l
		\nonumber\\
		&=&{\bf R}_1^T(\theta_1^l){\bf R}_2^T(\theta_2^l)\Array{ccc}{0 && \\ & 0& \\ && 1}{\bf R}_2(\theta_2^l){\bf R}_1(\theta_1^l)
	\label{eq.massreconstruction}
	\end{eqnarray}
The form of the mass matrix is determined by only two parameters $\theta_1^l$ and $\theta_2^l$. It is shown that ${\bf R}_3$ rotation angles $\theta^l_3$ are redundant for mass matrix ${\bf M}_0^l$. Therefore, rotation angles $\theta^\nu_3$ and $\theta^e_3$ are free parameters for lepton flavor mixing. In fact, it is just a result of $SO(2)_L^e\times SO(2)_L^\nu$ symmetry in the hierarchy limit. 
From eq. (\ref{eq.massreconstruction}), rotation angles $\theta_1^e,\theta_2^e,\theta_1^\nu,\theta_2^\nu$ are determined by the charged lepton and neutrino mass matrixes. However, 4 extra parameters, $\theta_3^e$, $\theta^\nu_3$, $\lambda^\nu_1$ and $\lambda_2^\nu$, correspond to PMNS mixing, i.e., ${\bf U}_{PMNS}={\bf U}_{PMNS}(\theta_3^e,\theta_3^\nu, \lambda_1^\nu,\lambda_2^\nu)$. 
This means that fermion hierarchy and flavor mixing have been divided into two independent questions controlled by two sets of parameters. 
The result provides a clear flavor structure without any nonphysical parameters.

%%%%
\section{Mass Matrix Reconstruction}

In terms of eq. (\ref{eq.massreconstruction}), the mass matrix ${\bf M}_0^{\nu,e}$ can be reconstructed from rotation angles $\theta_{1,2}^{\nu, e}$ on the Yukawa basis. Because $({\bf K}_L^e)^\dag {\bf U}_{PMNS}{\bf K}_L^\nu$ with
${\bf K}^{e}_L={\rm diag}(1,e^{i\beta^{e}_2},e^{i\beta^e_3}),
{\bf K}^{\nu}_L={\rm diag}(e^{i\beta_{1}^{\nu}},e^{i\beta^{\nu}_2},e^{i\beta^{\nu}_3})$ has the same PMNS mixings results as ${\bf U}_{PMNS}$, we must first add the diagonal matrix into ${\bf U}_{PMNS}$ to complement the nonphysical phases eliminated by rephasing.
Rotation transformation diagonalizes $({\bf K}_L^e)^\dag {\bf U}_{PMNS}{\bf K}_L^\nu$ into unitary eigenvalues
	\begin{eqnarray}
		{\bf R}_1^T(\theta_1^e){\bf R}_2^T(\theta_2^e){\bf R}_3^T(\theta_3^e)({\bf K}_L^e)^\dag {\bf U}_{PMNS}{\bf K}_L^\nu{\bf R}_3(\theta_3^\nu){\bf R}_2(\theta_2^\nu){\bf R}_1(\theta_1^\nu)
		=e^{-i\lambda_0^\nu}{\rm diag}(1, e^{-i\lambda^\nu_1},e^{-i\lambda_2^\nu})
		\label{eq.solve6theta}
	\end{eqnarray}
For a set of random initial ${\bf K}^{\nu,e}_L$, six rotation angles are solved from six non-diagonal equations.
We scanned the whole range of ${\bf K}^{\nu,e}_L$ and obtained all possible rotation angles $\theta_i^{\nu,e}$.
By analyzing the mixing angles, some characteristics in lepton mixings may be revealed. 

As we have shown in eq. (\ref{eq.II05}), complex phases required by nonvanishing CPV in ${\bf M}^l$ have been separated on the Yukawa basis. 
In the hierarchy limit, the pattern of ${\bf M}_0^l$ is determined by rotation angles $\theta_{1,2}^l$ in Eq. (\ref{eq.massreconstruction}) that is encoded into PMNS mixing matrix in terms of Eq. (\ref{eq.III07}).
Now, we are particularly interested in a common mass structure of up-type and down-type leptons.
If neutrinos and charged leptons obtain their hierarchical masses from a common mechanism, the question should be responded to by the similar and even the same mass matrix in the hierarchy limit. 
The answer can be found from the results of $\theta_1^l$ and $\theta_2^l$ in eq. (\ref{eq.solve6theta}).

A deviation degree between normalized up- and down-type lepton mass matrixes is defined as
	\begin{eqnarray}
		d_M= \frac{\sum_{i,j}\Big|(\frac{1}{\sum_km_k^\nu}{\bf M}_0^\nu-\frac{1}{\sum_km_k^e}{\bf M}_0^e)_{ij}\Big|^2}{\sum_{i,j}\Big|(\frac{1}{\sum_k m_k^\nu}{\bf M}_0^\nu+\frac{1}{\sum_km_k^e}{\bf M}_0^e)_{ij}\Big|^2}
	\label{eq.chifunction}
	\end{eqnarray}
In terms of eq. (\ref{eq.massreconstruction}), $d_M$ is determined by 4 rotation angles $\theta_{1}^{\nu},\theta_{2}^{\nu},\theta_{1}^{e},\theta_{2}^{e}$. 
Obviously, for the case $\theta_1^\nu-\theta_1^e=\theta_2^\nu-\theta_2^e=0$, ${\bf M}_0^\nu$ and ${\bf M}_0^e$ have the same structure, and $d_M$ takes a minimum value $d_M=0$.

\begin{figure}\caption{Evolution of $d_M$. Generation number $N=1,60,200,800$, respectively.}
		\centering
			\includegraphics[height=0.21 \textheight]{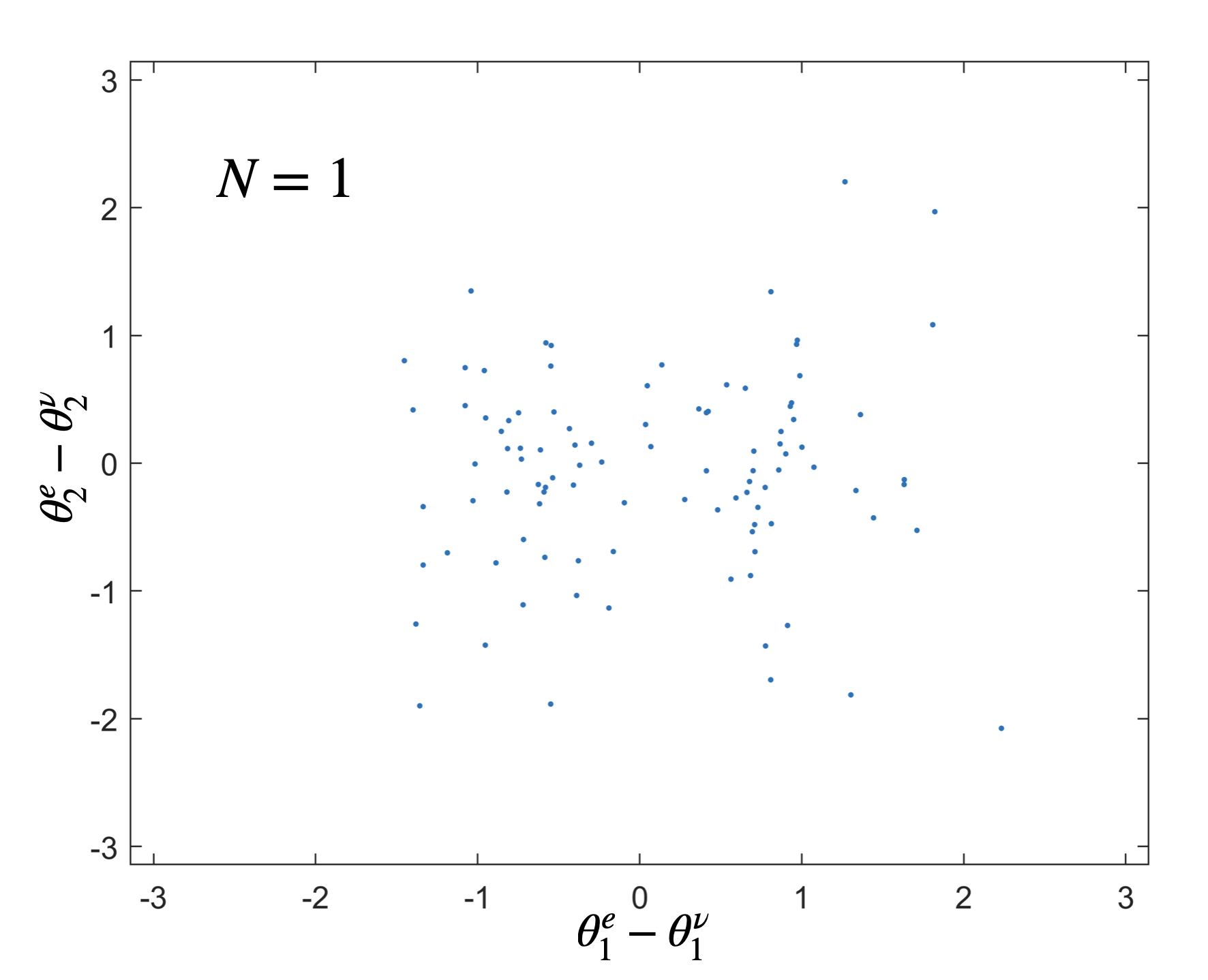}  
			\includegraphics[height=0.21 \textheight]{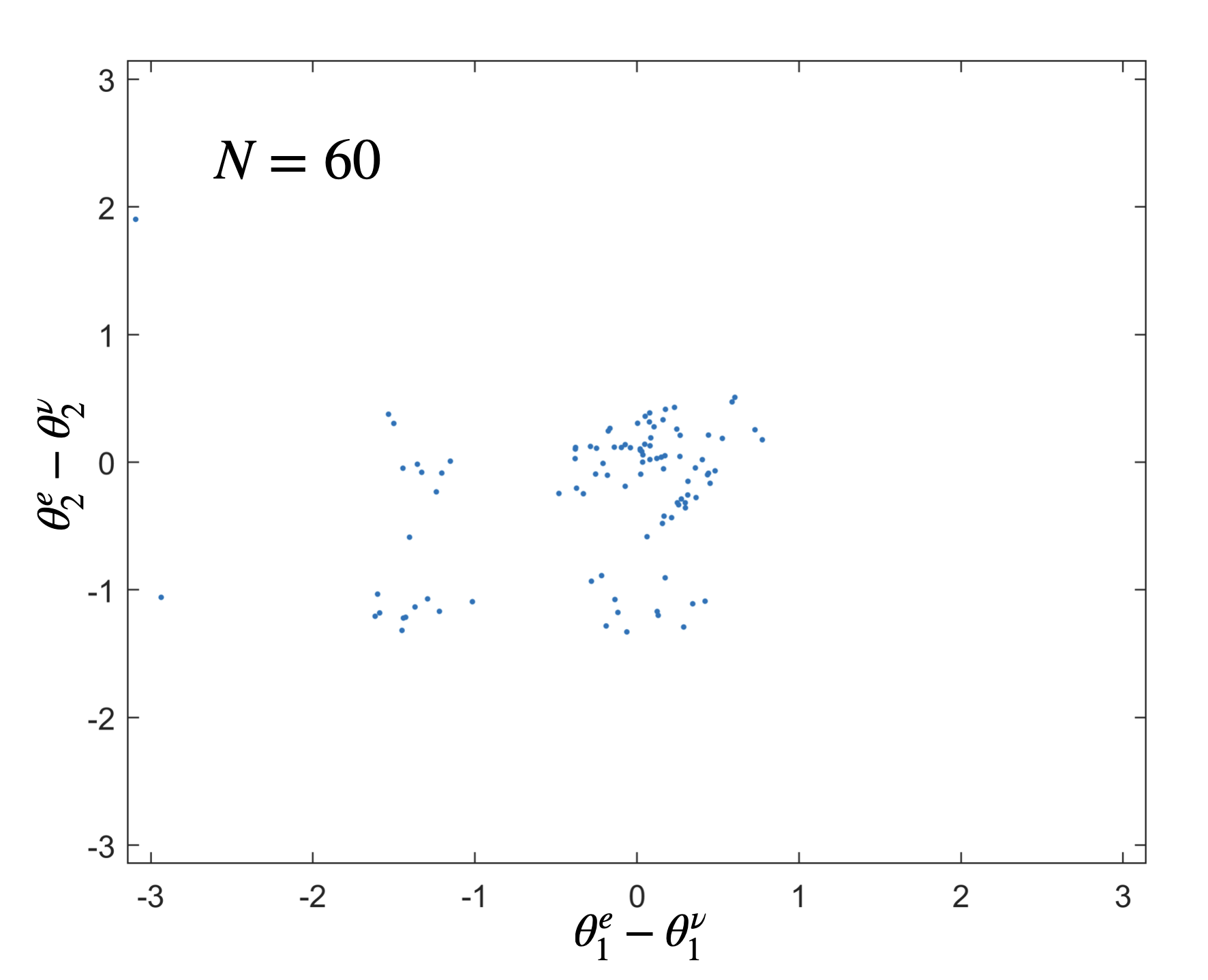}  
			\\
			\includegraphics[height=0.21 \textheight]{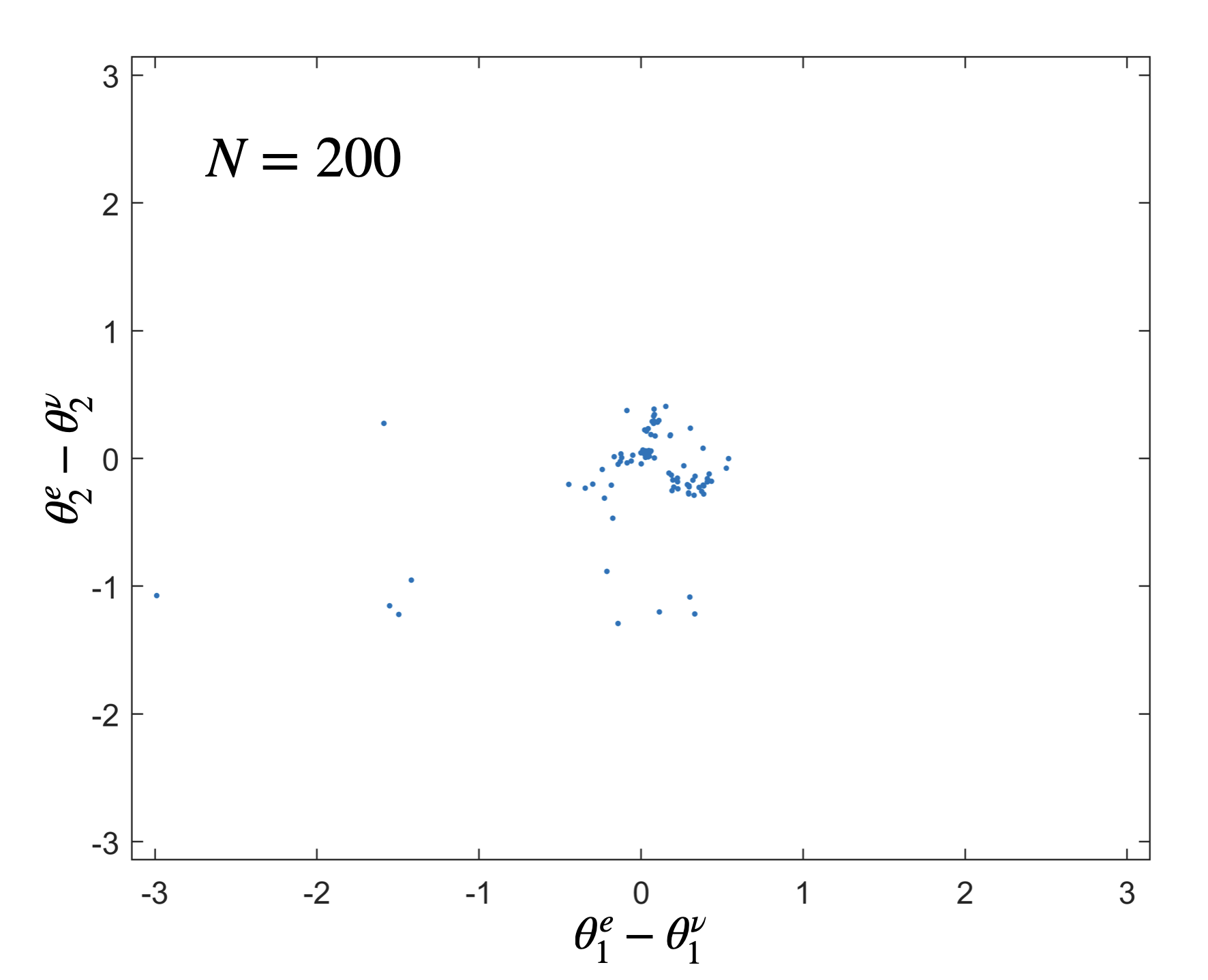}  
			\includegraphics[height=0.21 \textheight]{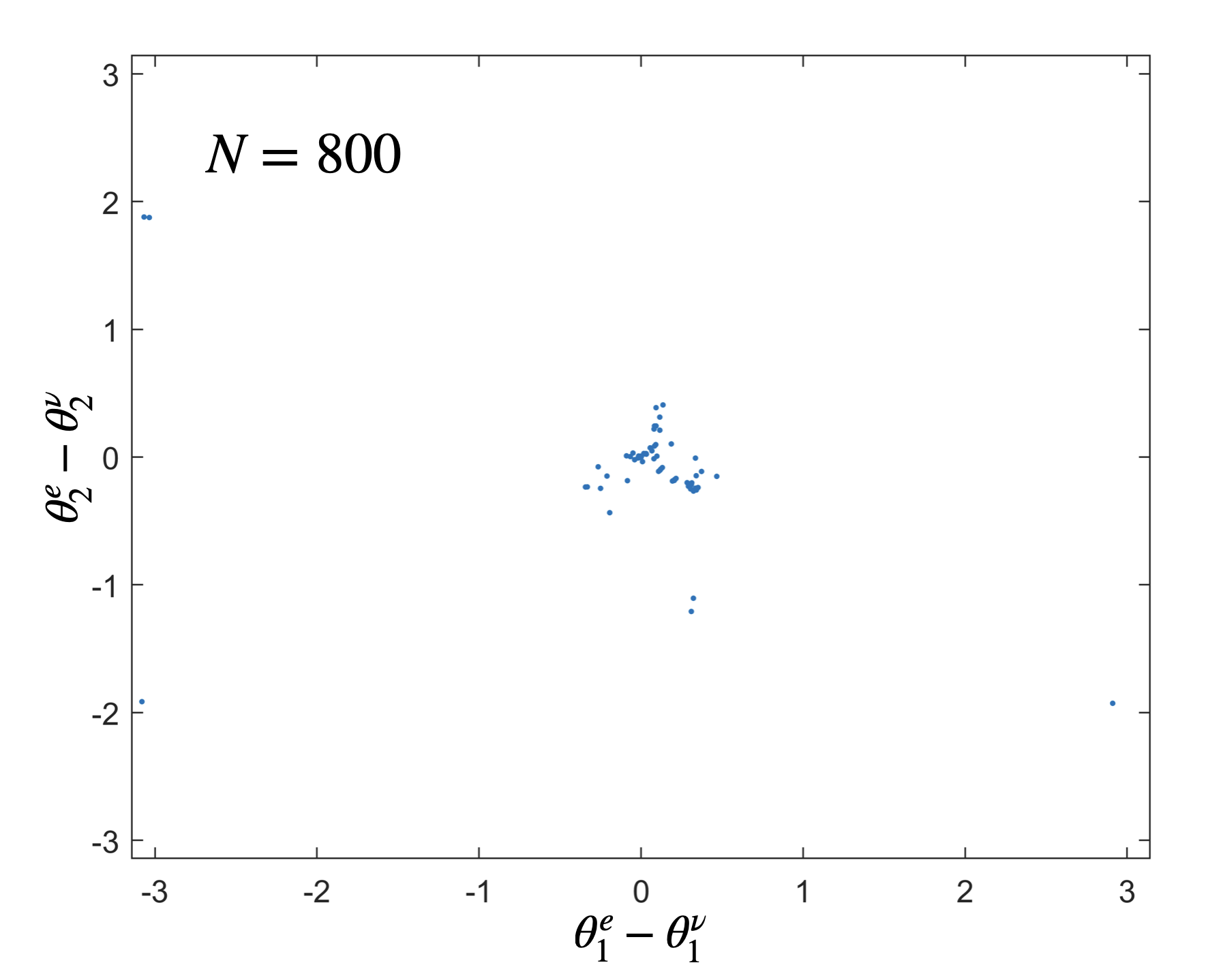}  
		\label{chisqevolution}
		\end{figure}

To find the minimum value of $d_M$, a differential evolution algorithm inspired by biological evolution is adopted as an iterative numerical technique. 
First, we initialize five random phases in ${\bf K}_{L}^{e,\nu}$ and get a genome from them. Genome means information of five arranged phases.
A value of $d_M$ can be calculated for each genome in terms of eq. (\ref{eq.chifunction}).
We initialize 100 genomes in the first generation and then choose the first genome to recombine with a random genome, among other genomes. If $d_M$ is decreased, we will replace the former one with this new genome. If not, the former one remains. We repeat this process with the next genome until all 100 genomes have evolved. To avoid falling into a local minimum, we introduce variation by allowing each individual to gain a randomly arranged variation, i.e. every number of a genome has a probability to mutate to a new genome. In physics, genome mutation means that a set of new phase parameters is generated near the original phases. By genome mutation, the parameter space is enlarged.
In the variation, we keep the better one as well. In the end, we obtain 100 new genomes and then put them into a new round to let the fittest genome emerge. Here, the mutation rates are set as $0.004\sim0.6$.
In the evolution procedure, genome modulation is also considered by linear recombination of two genome $G_1,G_2$ codes to a new one as $(1-\kappa)G_1+\kappa G_2$.
Here, the modulation factor $\kappa$ is set as $0.2$. 
From the viewpoint of the algorithm, modulation yields a set of new parameters between two original ones.
After $800$ generations, the minimal value of $d_M$ for the fittest genome is $\red{0.0004348}$.
 The results are plotted in the plane of $(\theta_1^\nu-\theta_1^e)$-$(\theta_2^\nu-\theta_2^e)$ in Fig.\ref{chisqevolution}. 
It has been shown that the trend of evolution of genomes approaches the origin in the plane, which is the position of the minimum $d_M$.
The results are listed as follows:
\red{
	\begin{eqnarray}
		\theta_1^\nu={-0.7844},~~~
		\theta_2^\nu={-0.6154},~~~
		\theta_3^\nu={-0.0786},
		\\
		\theta_1^e={-0.7853},~~~
		\theta_2^e={-0.6155},~~~
		\theta_3^e={0.4396},
		\\
		\lambda_0^\nu=2.075,~~~
		\lambda_1^\nu=-4.344,~~~
		\lambda_2^\nu=0.2510,
		\\
		\beta_2^e=1.304,~~~
		\beta_3^e=-0.6728,
		\\
		\beta_1^\nu=4.101,~~~
		\beta_2^\nu=-2.221,~~~
		\beta_3^\nu=-3.380.
	\end{eqnarray}
}
The corresponding PMNS input in eq. (\ref{eq.solve6theta}) is listed in Tab. (\ref{tab.PMNSresult}).
	\begin{table}[htp]
	\caption{PMNS mixing data from ref. \cite{2022PDG}.}
	\begin{center}
	\begin{tabular}{c|c|c|c|c}
	\hline\hline
		para. & $s^2_{12}$  & $s^2_{23}$ & $s^2_{13}$ & $\delta_{CP}$
	\\
	\hline
		input & \red{$0.307$} & \red{$0.547$} & \red{$0.0220$} & \red{$1.23\pi$}
	\\
	\hline
	\end{tabular}
	\end{center}
	\label{tab.PMNSresult}
	\end{table}%

Using eq. (\ref{eq.massreconstruction}), normalized  ${\bf M}_{0}^{\nu,e}$  are reconstructed as
	\begin{eqnarray}
		{\bf M}_{0}^{\nu}=\frac{1}{3}\sum_im_i^\nu
		\red{\Array{ccc}{0.9999 & 0.9990 & 1.0009
		\\ 
		0.9990 & 0.9981 & 1.0000
		\\ 
		1.0009 & 1.0000 & 1.0020}}
		\\
		{\bf M}_{0}^{e}=\frac{1}{3}\sum_im_i^e
		\red{\Array{ccc}{1.0002 & 1.0000 &1.0001
		\\ 
		1.0000 & 0.9998 & 0.9999
		\\
		1.0001 & 0.9999 &1.0000}}
	\end{eqnarray}
The above results strongly imply a flat matrix with all elements equal to $1$ as a common mass structure of charged leptons and neutrinos. This similar matrix has long been used as a democratic matrix in neutrino physics and the quark sector \cite{Sogami1998,Miura2000}. There are some essential differences: (1) the democratic matrix was proposed from discrete flavor symmetry based on some PMNS mixing modes, or it was assumed directly in the quark sector; (2) to address mass hierarchy as well as CPV in neutrino mixings, the democratic matrix must be corrected by some complex matrix elements close to 1, not real ones; and (3) the democratic matrix is only used as a neutrino mass structure in which the charged lepton mass matrix keeps the diagonal mass basis (or approximately).

Now, the flat structure of both neutrinos and charged leptons naturally arises from the homology of their masses. More importantly, the flat structure is not relative to any characteristic or symmetry in PMNS because flavor mixing has been isolated from the flavor structure.

To study the common mass structure, it is assumed that the lepton mass matrix in the hierarchy limit is addressed from the flat matrix ${\bf I}_0$:
	\begin{eqnarray}
		{\bf I}_0\equiv\Array{ccc}{1&1&1\\1&1&1\\1&1&1}=\frac{3}{\sum_im_i^l}{\bf M}_{0}^l
	\end{eqnarray}
It can be diagonalized by an orthogonal rotation $S_0$:
	\begin{eqnarray}
	{\bf S}_0{\bf I}_0{\bf S}_0^T ={\rm diag}(0,0,3)
	\end{eqnarray}
with
	\begin{eqnarray}
		{\bf S}_0=\frac{1}{\sqrt{6}}\Array{ccc}{\sqrt{3}& 0&-\sqrt{3}\\ -1& 2 & -1 \\ \sqrt{2} & \sqrt{2} & \sqrt{2}}
	\end{eqnarray}
There are two independent $SO(2)_L^l\times SO(2)_R^l$ symmetries corresponding to ${\bf R}_3(\theta_L^l)$ and ${\bf R}_3(\theta_R^l)$ for left-handed and right-handed leptons, respectively:
	\begin{eqnarray}
		\frac{1}{\sum_im_i^l}{\bf R}_3(\theta_L^l){\bf S}_0 {\bf M}_0^l{\bf S}_0^T{\bf R}_3^T(\theta_R^l)={\rm diag}(0,0,1)
	\label{eq.so2diagonalize}
	\end{eqnarray}
A similar transformation matrix was also discussed in \cite{Fritzsch2017CPC}, which is different from $S_0$ by a $SO(2)$ similarity transformation.
Now, the structure of PMNS is clearly expressed as
	\begin{eqnarray}
		{{\bf U}}_{PMNS}={\bf R}_3(\theta^e){\bf S}_0{\rm diag}(1, e^{-i\lambda_1^\nu},e^{-i\lambda_2^\nu}){\bf S}_0^T[{\bf R}_3(\theta^\nu)]^T
		\label{eq.PMNSs0}
	\end{eqnarray}
Here, the unphysical global phase $e^{-i\lambda_0}$ has been eliminated by rephasing. The structure of the lepton mixing matrix only includes the same number of parameters as the number of observables. Furthermore, these parameters are independent of the structure of the mass matrix. The mixing angles are determined from 
	\begin{eqnarray}
		s_{13}&=&|{\bf U}_{{PMNS},13}|
		\\
		s_{12}^2&=&\frac{|{\bf U}_{PMNS,12}|^2}{1-|{\bf U}_{PMNS,13}|^2}
		\\
		s_{23}^2&=&\frac{|{\bf U}_{PMNS,23}|^2}{1-|{\bf U}_{PMNS,13}|^2}
	\end{eqnarray}
and the CP violation phase
	\begin{eqnarray}
		c_{12}^2c_{23}^2+s_{13}^2s_{23}^2s_{12}^2-2c_{12}s_{12}c_{23}s_{23}s_{13}\cos[\delta_{CP}]
		=|{\bf U}_{PMNS,,22}|^2
\end{eqnarray}

\section{from PMNS to CKM}
By replacing neutrinos and charged leptons with up-type and down-type quarks, respectively, all of the above results can be generalized to the quark sector. Quarks in the Yukawa basis are defined by
	\begin{eqnarray*}
		u^{(Y)}_{L,R}={\bf F}_{L,R}^u u_{L,R},~~~
		d^{(Y)}_{L,R}={\bf F}_{L,R}^d d_{L,R}
	\end{eqnarray*}
On this basis, the quark mass matrix obtains the real form ${\bf M}_0^{q}$ (for $q=u,d$) and can be diagonalized by real rotation ${\bf U}_0^q$ as
	\begin{eqnarray*}
		{\bf U}_0^q{\bf M}_0^q({\bf U}_0^q)^T={\rm diag}(m_1^q,m_2^q,m_3^q)
	\end{eqnarray*}
The minimal parameterized ${\bf F}_{L,R}^{u,d}$ is assigned with only two relative phases for left-handed quarks as
	\begin{eqnarray*}
		{\bf F}_L^u={\rm diag}(1,e^{i\lambda_1^u},e^{i\lambda_2^u}),~~
		{\bf F}_L^d={\bf F}_R^u={\bf F}_R^d={\rm diag}(1,1,1)
	\end{eqnarray*}
The CKM matrix that appears as a quark weakly charged current term is expressed as
	\begin{eqnarray}
		{\bf U}_{CKM}={\bf U}_0^u{\rm diag}(1,e^{i\lambda_1^u},e^{i\lambda_2^u})({\bf U}_0^d)^T
		\label{eq.CKM001}
	\end{eqnarray}
Due to the similar quark hierarchal structure as a lepton, there is also an approximate $SO(2)_L^u\times SO(2)_R^u\times SO(2)_L^d\times SO(2)_R^d$ symmetry in the quark hierarchy limit. The general CKM mixing matrix is written as
\begin{eqnarray}
	{\bf U}_{CKM}=  e^{i\lambda_0^u}{\bf R}_3(\theta_3^u){\bf R}_2(\theta_2^u){\bf R}_1(\theta_1^u)\text{~diag}\left(1,e^{i\lambda^u_1},e^{i\lambda^u_2}\right){\bf R}_1^T(\theta_1^d){\bf R}_2^T(\theta_2^d){\bf R}_3^T(\theta_3^d)
	\label{eq.CKM37}
	\end{eqnarray}
In the hierarchy limit, $SO(2)_L^u$ and $SO(2)_L^d$ provide two free rotations ${\bf R}_3(\theta_L^{u})$ and ${\bf R}_3(\theta_L^d)$ to CKM, similar to eq. (\ref{eq.PMNSstructure01}) for PMNS.
The other rotation angles $\theta_{1}^{u,d}, \theta_{2}^{u,d}$ determine the normalized quark mass matrix by	\begin{eqnarray}
		\frac{1}{\sum_im_i^q}{\bf M}_0^q={\bf R}_1^T(\theta_1^q){\bf R}_2^T(\theta_2^q)\Array{ccc}{0 && \\ & 0& \\ && 1}{\bf R}_2(\theta_2^q){\bf R}_1(\theta_1^q)
	\label{eq.massreconstruction02}
	\end{eqnarray}
Inputting the ${\bf U}_{CKM}$ experimental data listed in Tab. \ref{tab.CKMresult}, all possible rotation angles $\theta_i^{u,d}$ are solved. Under similar attention to the homology of quark mass, we obtain a result almost the same as $\theta_i^u$ and $\theta_i^d$ for $i=1,2$.
The results are listed as follows:
\red{
\begin{eqnarray}
		\theta_1^u=-0.7807,~~~
		\theta_2^u=-0.6178,~~~
		\theta_3^u=0.1166,
		\\
		\theta_1^d=-0.7900,~~~
		\theta_2^d=-0.6136,~~~
		\theta_3^d=-0.1.32,
		\\
		\lambda_0^u=-2.0924,~~~
		\lambda_1^u=0.0551,~~~
		\lambda_2^u=-0.0454
	\end{eqnarray}
}

	\begin{table}[htp]
	\caption{CKM mixing data from ref. \cite{2022PDG}. Alternatively, the Wolfenstein parameters are defined in \cite{WolfensteinPRL1983}}
	\begin{center}
	\begin{tabular}{c|c|c|c|c}
	\hline\hline
		para. & $s_{12}$  & $s_{23}$ & $s_{13}$ & $\delta_{CP}$
	\\
	\hline
		input & \red{$0.22500$} & \red{$0.04182$} & \red{$0.00369$} & \red{$1.144$} 
	\\
	\hline
	\end{tabular}
	\end{center}
	\label{tab.CKMresult}
	\end{table}%
Using eq. (\ref{eq.massreconstruction02}),  ${\bf M}_{0}^{u,d}$ are reconstructed as
	\begin{eqnarray}
		{\bf M}_{0}^{u}=\frac{1}{3}\sum_im_i^u
		\red{\Array{ccc}{
			1.0065&0.9969&1.0063
			\\
			0.9969&0.9873&0.9967
			\\
			1.0063&0.9967&1.0061}},~
		{\bf M}_{0}^{d}=\frac{1}{3}\sum_im_i^d
		\red{\Array{ccc}{
			0.9946&1.0032&0.9940
			\\ 
			1.0032&1.0120&1.0027
			\\ 
			0.9940&1.0027&0.9935
			}}
	\end{eqnarray}
The flat matrix again appears as a common mass structure of quarks. 
Now, the homology of mass is generalized from not only up-type and down-type fermions but also all families of leptons and quarks. It is a natural characteristic of a fermion mass hierarchy structure.

According to the flat structure of the quark mass matrix, the CKM mixing matrix is also expressed by
	\begin{eqnarray}
		{{\bf U}}_{CKM}={\bf R}_3(\theta^u){\bf S}_0{\rm diag}(1, e^{i\lambda_1^u},e^{i\lambda_2^u}){\bf S}_0^T[{\bf R}_3(\theta^d)]^T
	\end{eqnarray}
%%%%

\section{Inspiration from the Flat Matrix}
The masses of all fermion families have pointed to the same flat structure together. 
It is a clue to the fermion mass generation. As an example, the quark mass term is rewritten on the Yukawa basis as
	\begin{eqnarray}
		-\mathcal{L}_M^q&=&\bar{u}_L^{(Y)}{\bf M}_0^u u_R^{(Y)}+\bar{d}_L^{(Y)}{\bf M}_0^d d_R^{(Y)}+h.c.
		\nonumber\\
		&=&\frac{\sum_im_i^u}{3}\bar{u}_L^{(Y)}{\bf I}_0 u_R^{(Y)}+\frac{\sum_im_i^d}{3}\bar{d}_L^{(Y)}{\bf I}_0 d_R^{(Y)}+h.c.
		\nonumber\\
		&=&\frac{\sum_im_i^u}{3}\Big(\bar{u}_{L,1}^{(Y)}+\bar{u}_{L,2}^{(Y)}+\bar{u}_{L,3}^{(Y)}\Big)\Big(u_{R,1}^{(Y)}+u_{R,2}^{(Y)}+u_{R,3}^{(Y)}\Big)+\Big(u\rightarrow d\Big)+h.c.
		\label{eq.Flat01}
	\end{eqnarray}
This means that the family mass $\sum_i m_i^{q}$ is generated as a whole. There exists an undifferentiated mass element $\bar{q}_{L,i}^{(Y)}q_{R,j}^{(Y)}$ between any two generations in a family. If the masses are generated from the Higgs mechanism, as in the SM, the above mass term suggests a family-universal Yukawa coupling $y_q$ as follows:
	\begin{eqnarray}
		-\mathcal{L}_M^q
		&=&y^u\Big(\sum_i\bar{Q}_{L,i}^{(Y)}\Big)\tilde{H}\Big(\sum_ju_{R,j}^{(Y)}\Big)+y^d\Big(\sum_i\bar{Q}_{L,i}^{(Y)}\Big)H\Big(\sum_jd_{R,j}^{(Y)}\Big)+h.c.
		\label{eq.Flat02}
	\end{eqnarray}
with the total family mass $\sum_im_i^q/3=y^qv_0/\sqrt{2}$. 

Essentially, the flat flavor structure arises from hierarchal masses of fermions in a family. The structure may be generalized to up-type and down-type fermions inspired by the hierarchal family mass $\sum_i m_i^u\gg \sum_i m_i^d$ in the quark sector and $\sum_i m_i^e\gg \sum_i m_i^\nu$ in the lepton sector. Even flat structures also play the same role between quarks and leptons. This motivation has been studied in \cite{YZhang202102}. The results showed that family-universal Yukawa coupling could be unified into quark/lepton coupling and even universal coupling for all fermions, such as gauge couplings in the GUT.

The remaining question is how to break flavor. Because flavor mixing and mass hierarchy are two independent problems, flavor breaking must be addressed from the corrections of the flat mass matrix. 
In \cite{Fritzsch2017CPC}, the flavor breaking of democracy of quark flavors was discussed in parallel with approximate CKM mixing values.
In phenomenology, flavor breaking requires two independent parameters corresponding to different $h_{12}^f$ and $h_{23}^f$ in a family.
In our early research \cite{YZhang202101}, flavor breaking was suggested by symmetric real corrections in nondiagonal elements of ${\bf I}_0$:
	\begin{eqnarray}
		{\bf I}_0\xrightarrow{flavor~breaking}{\bf I}_\Delta^f=\Array{ccc}{1& 1+\delta_{12}^f & 1+\delta_{13}^f \\ 1+\delta_{12}^f & 1 & 1+\delta_{23}^f \\ 1+\delta_{13}^f & 1+\delta_{23}^f & 1}.
	\end{eqnarray}
The superscript $^f$ labels flavor-dependent characteristics. The mass eigenvalues with flavor breaking corrections are
\begin{eqnarray}
		m^f_{1,2}&=&\frac{y^fv_0}{\sqrt{2}}\left(\frac{1}{3}S^f\mp\frac{2}{3}\sqrt{Q^f}\right)+\mathcal{O}(\delta^2)
		\label{eq.m1m2}\\
		m^f_3&=&\frac{y^fv_0}{\sqrt{2}}\left(3-\frac{2}{3}S^f\right)+\mathcal{O}(\delta^2)
		 \label{eq.m3}
	\end{eqnarray}
with parameters $S^f,Q^f$ as
	\begin{eqnarray}
		S^f&\equiv& -\delta^f_{12}-\delta^f_{23}-\delta^f_{13}
\label{eq.S}\\
		Q^f&\equiv&(\delta^f_{12})^2+(\delta^f_{23})^2+(\delta^f_{13})^2-\delta^f_{12}\delta^f_{23}-\delta^f_{23}\delta^f_{13}-\delta^f_{13}\delta^f_{12}
\label{eq.Q}
	\end{eqnarray}
The two $S^f$ and $Q^f$ are determined by mass hierarchies $h_{12}^f$ and $h_{23}^f$:
\begin{eqnarray}
		S^f&=&\frac{9}{2}(h^f_{23}+h^f_{12}h^f_{23}-(h^f_{23})^2)+\mathcal{O}(h^3)
		\label{eq.hierarchy2SQ1}\\
		Q^f&=&\frac{81}{16}(h^f_{23})^2+\mathcal{O}(h^3)
	\label{eq.hierarchy2SQ2}
\end{eqnarray}
The remaining correction of the three $\delta_{ij}^f$ was shown as a $SO(2)_L^f$ rotation angle. A prominent advantage of flavor breaking is that it provides a general way for the quark sector and lepton sector. More importantly, it was shown that this was an efficient way to fit quark/lepton hierarchal masses and CKM/PMNS with arbitrary precision.

\section{summary}
The Yukawa terms of the SM have been rewritten on a new basis. When expressing the flavor structure back into a mass basis or gauge basis, it is completely coincident with the SM after EW breaking. There is no new phenomenological prediction beyond the SM. However, the flavor structure of the SM fermions in the new Yukawa basis is shown clearly, and all nonphysical parameters in $y_{ij}^f$ are eliminated. 
The study's main results show that quark/lepton flavor mixing can be parameterized independently. In the hierarchy limit, CKM/PMNS is successfully expressed by two $[SO(2)_L]^2$ rotation angles and two relative phases. 
Using the decomposition of the flavor mixing matrix, up-type, and down-type fermion mass matrices are reconstructed. The results show that the homology of the fermion mass matrix points to a common flat mass structure for not only leptons but also quarks. We also discussed the inspiration of the flat mass structure on the unification of Yukawa interactions between different families and flavor breaking by nondiagonal corrections. 
Flavor breaking is still an open unknown question. More importantly, the results in the paper can help to understand the truth of flavor mixing and mass hierarchy more clearly and profoundly.

\section*{Acknowledgements}
I thank  Prof. Shao-zhou Jiang and Prof. Rong Li for their discussions on the subject. This work is supported by the Fundamental Research Funds for the Central Universities and by Shaanxi Natural Science Foundation (2022JM-052) and SAFS 22JSY035 of China..

\end{document}